# Close-Form Expression of One-Tap Normalized LMS Carrier Phase Recovery in Optical Communication Systems

Tianhua Xu[1,2,3,4,*], Gunnar Jacobsen[2,3], Sergei Popov[2], Jie Li[3], Tiegen Liu[4] and Yimo Zhang[4]
[1]University College London, London, WC1E7JE, United Kingdom
[2]Royal Institute of Technology, Stockholm, SE-16440, Sweden
[3]Acreo Swedish ICT AB, Stockholm, SE-16440, Sweden
[4]Tianjin University, Tianjin, 300072, China

(Invited)

## ABSTRACT

The performance of long-haul high speed coherent optical fiber communication systems is significantly degraded by the laser phase noise and the equalization enhanced phase noise (EEPN). In this paper, the analysis of the one-tap normalized least-mean-square (LMS) carrier phase recovery (CPR) is carried out and the close-form expression is investigated for quadrature phase shift keying (QPSK) coherent optical fiber communication systems, in compensating both laser phase noise and equalization enhanced phase noise. Numerical simulations have also been implemented to verify the theoretical analysis. It is found that the one-tap normalized least-mean-square algorithm gives the same analytical expression for predicting CPR bit-error-rate (BER) floors as the traditional differential carrier phase recovery, when both the laser phase noise and the equalization enhanced phase noise are taken into account.

## 1. INTRODUCTION

High speed optical fiber communication systems can be significantly deteriorated by system impairments, such as chromatic dispersion (CD), polarization mode dispersion (PMD), laser phase noise (PN) and fiber nonlinearities (FNLs) [1-4]. Coherent optical detection and digital signal processing (DSP) allow the powerful equalization and mitigation of the communication system impairments in the electrical domain, and have become one of the most promising techniques for the next-generation optical fiber transmission networks, with the full capture of the amplitude and the phase of the transmitted optical signals [5-12]. Some feed-forward and feed-back carrier phase estimation (CPE) algorithms have been proposed to compensate the phase noise from the laser sources [13-20]. Among these reported carrier phase recovery (CPR) methods, the one-tap normalized least-mean-square (LMS) equalizer has been validated for compensating the laser phase noise effectively in the high speed coherent optical fiber transmission systems [15-19].

In this paper, a theoretical assessment for the carrier phase recovery in the quadrature phase shift keying (QPSK) coherent optical transmission systems using the one-tap normalized LMS equalizer is discussed in detail. The analytical expression for this one-tap normalized LMS equalization has been presented in order to predict the bit-error-rate (BER) performance, e.g. the BER floor, in the carrier phase recovery process. It can be found that compared to the traditional differential carrier phase recovery, the one-tap normalized LMS equalization shows a similar performance for compensating the laser phase noise. The close-form prediction for the BER floor in the one-tap normalized LMS CPE algorithm gives the same expression as the differential carrier phase recovery.

Meanwhile, due to the interplay between the electronic dispersion equalization (EDC) and the laser phase noise, an effect of equalization enhanced phase noise (EEPN) has been generated and will seriously degrade the performance of long-haul optical fiber communication systems [21-30]. Considering the impact of EEPN, the traditional analysis of the carrier phase recovery is not appropriate any more for the design of the long-haul optical fiber transmission systems, e.g. the requirement on laser linewidths will not be relaxed with the increment of signal symbol rates. Therefore, it will also be interesting to investigate the BER performance in the one-tap normalized LMS carrier phase recovery, when the equalization enhanced phase noise is taken into account. This is also discussed and analyzed in detail in Section 3.

[*]tianhua.xu@ucl.ac.uk

## 2. PRINCIPLE OF LASER PHASE NOISE AND EQUALIZATION ENHANCED PHASE NOISE

In the QPSK coherent optical communication system, the variance of the phase noise from the transmitter (Tx) laser and the local oscillator (LO) laser can be expressed as follows, see e.g. [1]

$$\sigma^2_{Tx\_LO} = 2\pi(\Delta f_{Tx} + \Delta f_{LO})\,T_S \qquad (1)$$

where $\Delta f_{Tx}$ and $\Delta f_{LO}$ are the 3-dB linewidths of the Tx laser and the LO laser respectively, and $T_S$ is the symbol period of the transmission system. The noise variance of EEPN, due to the interaction between the dispersion and the LO laser phase noise in the EDC based optical transmission systems is expressed as follows, see e.g. [17,22]

$$\sigma^2_{EEPN\_LO} = \pi\lambda^2 DL\Delta f_{LO}/2cT_S \qquad (2)$$

where $f_{LO}$ is the LO laser central frequency, which is usually equal to the Tx laser central frequency $f_{Tx}$, $D$ is the CD coefficient of the transmission fiber, $L$ is the transmission fiber length, $R_S=1/T_S$ is the symbol rate of the transmission system, and $\lambda=c/f_{Tx}=c/f_{LO}$ is the central wavelength of the transmitted optical carrier wave.

When the effects from laser phase noise and the influence from EEPN are equal, we have $L_0 = 8cT_S^2/\lambda^2 D$. For 28-Gbaud dual-polarization QPSK coherent optical communication systems, $L_0$=79.27 km.

## 3. ANALYSIS OF NORMALIZED LMS CARRIER PHASE RECOVERY

### 3.1 Analysis of One-Tap Normalized LMS Equalizer

The transfer function of the one-tap normalized LMS carrier phase estimator can be expressed as follows,

$$y(n) = w(n)x(n) \qquad (3)$$

$$w(n+1) = w(n) + \mu e(n)x^*(n)/|x(n)|^2 \qquad (4)$$

$$e(n) = d(n) - y(n) \qquad (5)$$

We assume that the *n*-th input symbol is $x(n) = E_n \exp(j\varphi_n)$, and $\varphi_k$ is the carrier phase of the *n*-th input symbol. Then the *n*-th output symbol can be expressed as

$$y(n) = w(n)\cdot x(n) = b_n \exp(-j\Phi_n)\cdot E_n \exp(j\varphi_n) = b_n E_n \exp[j(\varphi_n - \Phi_n)] \qquad (6)$$

$$w(n) = b_n \exp(-j\Phi_n) \qquad (7)$$

where $\Phi_n$ is the estimated carrier phase, then phase error is $\Delta\Phi_n = \varphi_n - \Phi_n$, when $\Delta\Phi_n \approx 0$, we will have $|e(n)| \ll 1$, then the one-tap normalized LMS algorithm gets converged.

For (*n*+1)-th input symbol, we will have $x(n+1) = E_{n+1}\exp(j\varphi_{n+1})$, then the (*n*+1)-th output symbol will be

$$y(n+1) = w(n+1)x(n+1) = \left\{b_n \exp(-j\Phi_n) + \frac{\mu}{|x(n)|^2}e(n)\cdot[E_n\exp(j\varphi_n)]^*\right\}\cdot E_{n+1}\exp(j\varphi_{n+1}) \qquad (8)$$

$$\approx [b_n E_{n+1} + \mu e(n)E_n^* E_{n+1}/|E_k|^2]\cdot \exp[j(\varphi_{n+1} - \varphi_n)]$$

When $-\pi/4 < \varphi_{k+1} - \varphi_k < \pi/4$, the demodulation part will not cause any errors for QPSK coherent optical transmission systems. As we know, $\phi_{k+1} - \phi_k$ follows a Gaussian distribution of $f(x) = (1/\sqrt{2\pi}\sigma)\exp(-x^2/2\sigma^2)$, $\sigma^2 = 2\pi T_S \Delta f$ is the variance of the phase noise difference.

Therefore, the symbol-error-rate (SER) for the QPSK transmission systems can be calculated as follows,

$$P_{SER}(e) = \int_{-\infty}^{-\frac{\pi}{4}} f(x)dx + \int_{+\frac{\pi}{4}}^{+\infty} f(x)dx = 2\int_{+\frac{\pi}{4}}^{+\infty} \frac{1}{\sqrt{2\pi}\sigma} \exp\left(-\frac{x^2}{2\sigma^2}\right)dx = erfc\left(\frac{\pi}{4\sqrt{2}\sigma}\right) \quad (9)$$

Thus the BER floor induced by the one-tap normalized LMS carrier phase recovery can be derived accordingly,

$$BER_{floor} = \frac{1}{2}P_{SER}(e) = \frac{1}{2}erfc\left(\frac{\pi}{4\sqrt{2}\sigma}\right) \quad (10)$$

The close-form prediction for the BER floor in the one-tap normalized least-mean-square CPE algorithm gives the same expression as the differential carrier phase estimation. It means that the one-tap normalized least-mean-square equalization shows a very similar performance compared to the traditional differential carrier phase recovery [17,31,32].

### 3.2 Influence of EEPN in One-Tap Normalized LMS Equalization

When the EEPN is taken into account in the one-tap normalized least-mean-square carrier phase estimation, we have the total noise variance in the optical fiber transmission system as the following expression,

$$\sigma_{Total}^2 = \sigma_{Tx\_LO}^2 + \sigma_{EEPN}^2 = 2\pi T_S(\Delta f_{Tx} + \Delta f_{LO}) + \pi \lambda^2 D \cdot L \cdot \Delta f_{LO}/2cT_S \quad (11)$$

Therefore, the BER floor in the one-tap normalized least-mean-square carrier phase estimation considering the equalization enhanced phase noise can be evaluated as,

$$BER_{floor} = \frac{1}{2}erfc\left(\frac{\pi}{4\sqrt{2}\sigma_{Total}}\right) = \frac{1}{2}erfc\left(\frac{\pi}{4\sqrt{2}} \cdot \frac{1}{\sqrt{2\pi T_S(\Delta f_{Tx} + \Delta f_{LO}) + \pi \lambda^2 DL\Delta f_{LO}/2cT_S}}\right) \quad (12)$$

### 4. RESULTS

As illustrated in Fig. 1, the performance of the one-tap normalized LMS carrier phase recovery algorithm is investigated in the 28-Gbaud QPSK coherent optical transmission system, where both theoretical analysis and numerical simulations have been carried out. The standard single mode fiber (SSMF) with a CD coefficient of 16 ps/nm/km is employed in the analysis, and the attenuation, PMD, nonlinearities are neglected. It can be found in Fig. 1(a) that the performance the one-tap normalized LMS carrier phase recovery is degraded with the increment of transmission distance, and this effect is more serious for lager laser linewidths. In Fig. 1(b), it can be seen that the numerical simulation and the theoretical prediction achieve a good agreement, where the Tx and the LO linewidths are both 5 MHz and the transmission distance is 2000 km. This is a good verification for the proposed theoretical model of the one-tap normalized LMS carrier phase recovery.

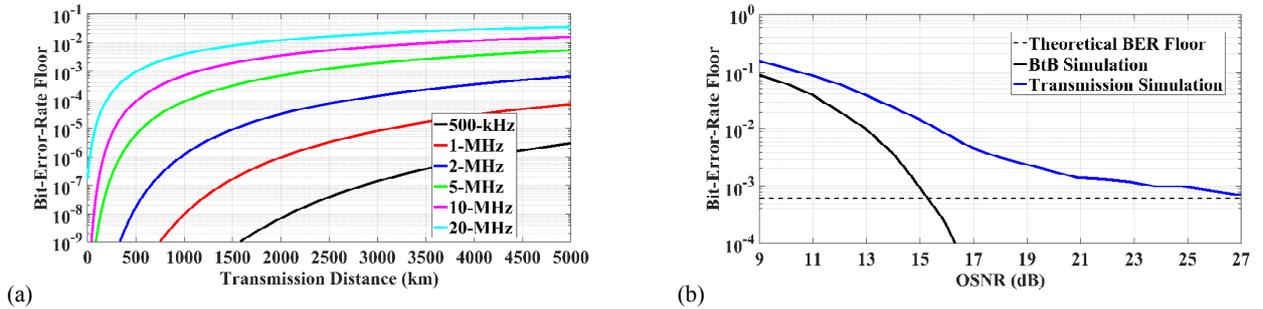

(a)  (b)

Fig. 1 Performance of the the one-tap normalized LMS carrier phase recovery in 28-Gbaud DP-QPSK transmission system (Tx laser linewidth is equal to LO laser linewidth).
(a) Theoretical BER floors for different transmission distances, (b) Comparison between numerical simulation and theoretical prediction

## 5. DISCUSSIONS

The derivation and the discussions for the one-tap normalized least-mean-square carrier phase recovery in the above were only carried out in the QPSK coherent optical transmission systems, for compensating both laser phase noise and equalization enhanced phase noise. However, these analyses can also be applied for *n*-level phase shift keying (*n*-PSK) coherent optical communication systems for both laser phase noise compensation and equalization enhanced phase noise compensation, by considering some reasonable effects and modifications in higher-level modulation formats. This analysis will be investigated in our future work and potential publications.

## 6. CONCLUSION

The theoretical evaluation of the one-tap normalized LMS carrier phase recovery in the long-haul high speed QPSK coherent optical communication systems has been investigated, both considering the laser phase noise and the equalization enhanced phase noise. It has been found that the one-tap normalized LMS carrier phase estimation shows a similar performance compared to the traditional differential carrier phase recovery in both cases.

## 7. ACKNOWLEDGMENT

This work is supported in parts by UK EPSRC project UNLOC EP/J017582/1, EU project GRIFFON 324391, EU project ICONE 608099, and Swedish Vetenskapsradet 0379801.

## REFERENCES

[1] G. P. Agrawal, Fiber-optic communication systems (4th Edition), John Wiley & Sons, New York (2010)
[2] T. Xu, et al., "Field trial over 820 km installed SSMF and its potential Terabit/s superchannel application with up to 57.5-Gbaud DP-QPSK transmission," Opt. Commun., **353**, 133-138, (2015)
[3] H. Zhang, et al., "A quantitative robustness evaluation model for optical fiber sensor networks," J. Lightwave Technol., **31**, 1240-1246, (2013)
[4] Y. Li, et al., "Dynamic dispersion compensation in a 40 Gb/s single-channeled optical fiber communication system," ACTA OPTICA SINICA, **27**, 1161-1165, (2007)
[5] T. Xu, et al., "Chromatic dispersion compensation in coherent transmission system using digital filters," Opt. Express, **18**, 16243-16257, (2010)
[6] S. J. Savory, "Digital filters for coherent optical receivers," Opt. Express, **16**, 804-817, (2008)
[7] T. Xu, et al., "Normalized LMS digital filter for chromatic dispersion equalization in 112-Gbit/s PDM-QPSK coherent optical transmission system," Opt. Commun., **283**, 963-967, (2010)
[8] R. Kudo, et al., "Coherent optical single carrier transmission using overlap frequency domain equalization for long-haul optical systems," J. Lightwave Technol., **27**, 3721-3728, (2009)
[9] T. Xu, et al., "Frequency-domain chromatic dispersion equalization using overlap-add methods in coherent optical system," J. Opt. Commun., **32**, 131-135, (2011)
[10] E. Ip and J. M. Kahn, "Digital equalization of chromatic dispersion and polarization mode dispersion," J. Lightwave Technol., 25, 2033-2043, (2007)
[11] G. Liga, et al., "On the performance of multichannel digital backpropagation in high-capacity long-haul optical transmission," Opt. Express, **22**, 30053-30062, (2014)
[12] R. Maher, et al., "Spectrally shaped DP-16QAM super-channel transmission with multi-channel digital back propagation," Sci. Rep., 5, 08214, (2015)
[13] G. Jacobsen, et al., "Receiver implemented RF pilot tone phase noise mitigation in coherent optical nPSK and nQAM systems," Opt. Express, **19**, 14487-14494, (2011)
[14] M. G. Taylor, "Phase estimation methods for optical coherent detection using digital signal processing," J. Lightwave Technol., **17**, 901-914, (2009)

[15] Y. Mori, et al., "Unrepeated 200-km transmission of 40-Gbit/s 16-QAM signals using digital coherent receiver," Opt. Express, **17**, 1435-1441, (2009)

[16] T. Xu, et al., "Carrier phase estimation methods in coherent transmission systems influenced by equalization enhanced phase noise," Opt. Commun., **293**, 54-60, (2013)

[17] T. Xu, et al., "Analytical estimation of phase noise influence in coherent transmission system with digital dispersion equalization," Opt. Express, **19**, 7756-7768, (2011)

[18] G. Jacobsen, T. Xu, S. Popov and S. Sergeyev, "Study of EEPN mitigation using modified RF pilot and Viterbi-Viterbi based phase noise compensation," Opt. Express, **21**, 12351-12362, (2013)

[19] T. Xu, et al., "Analysis of carrier phase extraction methods in 112-Gbit/s NRZ-PDM-QPSK coherent transmission system," Proc. Asia Commun. Photon. Conf., AS1C.2, (2012)

[20] A. J. Viterbi and A. M. Viterbi, "Nonlinear estimation of PSK-modulated carrier phase with application to burst digital transmission," IEEE Trans. Inf. Theory, **29**, 543-551, (1983)

[21] T. Xu, et al., "Equalization enhanced phase noise in Nyquist-spaced superchannel transmission systems using multi-channel digital back-propagation," Sci. Rep., **5**, 13990, (2015)

[22] W. Shieh and K. P. Ho, "Equalization-enhanced phase noise for coherent detection systems using electronic digital signal processing," Opt. Express, **16**, 15718-15727, (2008)

[23] C. Xie, "Local oscillator phase noise induced penalties in optical coherent detection systems using electronic chromatic dispersion compensation," Proc. Opt. Fiber Commun. Conf., OMT4, (2009)

[24] A. P. T. Lau, T. S. R. Shen, W. Shieh and K. P. Ho, "Equalization-enhanced phase noise for 100Gb/s transmission and beyond with coherent detection," Opt. Express, **18**, 17239-17251, (2010)

[25] G. Jacobsen, et al., "EEPN and CD study for coherent optical nPSK and nQAM systems with RF pilot based phase noise compensation," Opt. Express, **20**, 8862-8870, (2012)

[26] T. Xu, et al., "Analytical BER performance in differential n-PSK coherent transmission system influenced by equalization enhanced phase noise," Opt. Commun., **334**, 222-227, (2015)

[27] K. P. Ho and W. Shieh, "Equalization-enhanced phase noise in mode-division multiplexed systems," Journal of Lightwave Technology, Vol. 31(13), 2237-2243, 2013.

[28] I. Fatadin and S. J. Savory, "Impact of phase to amplitude noise conversion in coherent optical systems with digital dispersion compensation," Opt. Express, **18**, 16273-16278, (2010)

[29] G. Colavolpe, T. Foggi, E. Forestieri, M. Secondini, "Impact of phase noise and compensation techniques in coherent optical systems," Journal of Lightwave Technology, Vol. 29(18), 2790-2800, 2011.

[30] G. Jacobsen, et al., "Influence of pre- and post-compensation of chromatic dispersion on equalization enhanced phase noise in coherent multilevel systems," J. Opt. Commun., **32**, 257-261, (2011)

[31] J. G. Proakis and M. Salehi, Digital communications (5th Edition), McGraw-Hill, New York, (2007)

[32] I. Fatadin, D. Ives and S. J. Savory, "Differential carrier phase recovery for QPSK optical coherent systems with integrated tunable lasers," Opt. Express, **21**, 10166-10171, (2013)